\documentclass[twocolumn]{revtex4}
\usepackage[utf8]{inputenc}

\begin{document}

\title{Quantum hydrodynamics of the spinor Bose-Einstein condensate at non-zero temperatures}

\author{Pavel A. Andreev}
\email{andreevpa@physics.msu.ru}
\affiliation{Faculty of physics, Lomonosov Moscow State University, Moscow, Russian Federation, 119991.}
\affiliation{Peoples Friendship University of Russia (RUDN University), 6 Miklukho-Maklaya Street, Moscow, 117198, Russian Federation}

\author{I. N. Mosaki}
\affiliation{Faculty of physics, Lomonosov Moscow State University, Moscow, Russian Federation, 119991.}

\author{Mariya Iv. Trukhanova}
\email{trukhanova@physics.msu.ru}
\affiliation{Faculty of physics, Lomonosov Moscow State University, Moscow, Russian Federation, 119991.}
\affiliation{Russian Academy of Sciences, Nuclear Safety Institute (IBRAE), B. Tulskaya 52, Moscow, Russian Federation, 115191.}

\date{\today}

\begin{abstract}
Finite temperature hydrodynamic model is derived for the spin-1 ultracold bosons
by the many-particle quantum hydrodynamic method.
It is presented as the two fluid model of the BEC and normal fluid.
The linear and quadratic Zeeman effects are included.
Scalar and spin-spin like short-range interactions are considered in the first order by the interaction radius.
It is also represented as the set of two nonlinear Pauli equations.
The spectrum of the bulk collective excitations is considered for the ferromagnetic phase in the small temperature limit.
The spin wave is not affected by the presence of the small temperature in the described minimal coupling model,
where the thermal part of the spin-current of the normal fluid is neglected.
The two sound waves are affected by the spin evolution in the same way as the change of spectrum of the single sound wave in BEC,
where speed of sound is proportional to $g_{1}+g_{2}$
with $g_{i}$ are the interaction constants.
\end{abstract}

\pacs{03.75.Hh, 03.75.Kk, 67.85.Pq}
\keywords{BEC, hydrodynamics, finite temperatures, sound waves.}


\maketitle


\section{Introduction}

The superfluidity of the liquid helium below the $\lambda$ point is discovered in 1938.
This phenomenon is explained as the Bose-Einstein condensation of the part of atoms.
The liquid helium is the physical object with strong interaction
while the Bose-Einstein condensation is predicted for the ideal Bose gas.
The Bose-Einstein condensation of collections of weakly interacting atoms is achieved in 1995.
The dynamics of the weakly interacting spin-0 scalar Bose-Einstein condensates (BECs) is described by the Gross-Pitaevskii equation
which is the effective single-particle nonlinear Schrodinger equation \cite{Dalfovo RMP 99}.
In the following years the Bose-Einstein condensate of atoms being in several spin states \cite{Stamper-Kurn RMP 13}, \cite{Kawaguchi PR 12}.
Corresponding model in form of the nonlinear Pauli equation is suggested.

The Gross-Pitaevskii equation can be represented in the form of quantum hydrodynamic equations.
The set of hydrodynamic equations consists of the continuity and Euler equation for the potential velocity field for the spin-0 BECs.
The velocity field becomes nonpotential for the spin-1 BECs.
Moreover, additional hydrodynamic functions appear to describe the hydrodynamic properties of the spin-1 BECs.
They are the spin density and the nematic tensor density.
The nematic tensor is proportional to the anticommutator of the spin operators.
It is an independent function for the spin-1 BECs
(except the ferromagnetic phase, where it is reduced to the spin density).
In contrast, the product of the spin operators reduces to its first degree for the spin-1/2 particles.
Overall, the set of hydrodynamic equations is obtained in the regime of small collisions
\cite{Fujimoto PRA 13}, \cite{Szirmai PRA 12}, \cite{Kudo PRA 10}.

The Gross-Pitaevskii equation can be derived from the microscopic quantum mechanics represented in the second quantization form \cite{Alon arXiv 21},
which is one of realizations for the transition from the microscopic to the macroscopic models.
The many-particle quantum hydrodynamics presents another approach giving
the representation of the many-particle wave function via the set of hydrodynamic functions
\cite{Maksimov QHM 99}, \cite{MaksimovTMP 2001}, \cite{Andreev PRA08}, \cite{Andreev IJMP B 13}, \cite{Andreev 2001}, \cite{Andreev 2005}.
The density-functional theory for spin-0 bosons is considered including the GP equation \cite{Saarikoski RMP 10}.

Earlier steps in the theory of weakly interacting BECs includes the derivation of spectrum of collective excitations
called the Bogoliubov spectrum obtained in 1947 \cite{Bogoliubov 47}.
This theory is based on the second quantization method.
Effects of the small temperatures are included by Popov in 1965 \cite{Popov JETP 65}, \cite{Griffin PRB 96}, \cite{Zaremba PRA 98}.

Variety of waves phenomena is observed in BECs.
It includes the density waves \cite{Nakamura PRA 14}, \cite{Roy PRA 14}, \cite{Trukhanova PS 20}
solitons \cite{Andreev LP 19}, \cite{Andreev LP 21}, \cite{Andreev 2009},
vortices
\cite{Gautam PRA 14},
turbulence,
shock waves \cite{Crosta PRA 12}, \cite{Kamchatnov PRA 12}.
Collective excitations are considered in Fermi gas either \cite{Korolyuk PRA 14}.
These phenomena demonstrate more complex behavior in the spinor BECs in comparison with spin-0 BECs
\cite{Fujimoto PRA 13}.

Finite temperature effects in ultracold gases are considered as well
\cite{Roy PRA 14}, \cite{Gautam PRA 14}, \cite{Ticknor PRA 12}, \cite{Liu PRA 03}, \cite{Liu PRA 03 2}.
Presence of the bosons in the excited states is modeled via the second fluid which called the normal fluid.
The interaction
between the atoms in the condensate and the surrounding gas of noncondensed particles
can be also modeled within the kinetic model
\cite{Morandi PRA 13}.
Finite temperature effects are considered in ultracold fermions \cite{Korolyuk PRA 14} either.

Depletion of BECs and quantum fluctuations make
the dynamical properties bosons intermixing with the finite temperature effects more complex \cite{Andreev 2005},
\cite{Blakie PRA 13}, \cite{Pitaevskii PRL 98}, \cite{Braaten PRL 99}, \cite{Astrakharchik PRL 05}, \cite{Xu PRL 06}, \cite{Andreev 2101}.


mean-field potential of the magnetic dipole-dipole interaction is included at the study of the collective excitations in spin-1 BECs \cite{Deng PRA 20}.

Instantaneous quench of quadratic Zeeman shift from
$q_{1}>0$ to $q_{2}<0$ through the quantum phase transition at $q=0$ corresponding to the transition from polar to the antiferromagnetic phases is made experimentally
for the antiferromagnetic spinor BEC \cite{Vinit PRA 17}.



This paper is organized as follows.
In Sec. II the structure of the hydrodynamic equations including basic definitions and major steps of derivation are described.
In Sec. III transition to the two-fluid quantum hydrodynamic model for the finite temperature spin-1 bosons is made.
In Sec. IV the spectrum of collective excitations is considered for the ferromagnetic phase.
In Sec. V a brief summary of obtained results is presented.

\section{Hydrodynamic equations}

The dynamic of the quantum system is governed by the Schrodinger equation
\begin{equation}\label{BECfinTSpin1 Schrodinger equation micro} \imath\hbar\partial_{t}\Psi_{S}=\hat{H}_{SS'}\Psi_{S'}, \end{equation}
where the coordinate representation is chosen.
Therefore, the wave function $\Psi_{S}=\Psi_{S}(R,t)$ is the function of coordinates of $N$ identical particles under consideration
$R=\{\textbf{r}_{1}, ..., \textbf{r}_{N}\}$ representing the configurational space of the system.
The spin-1 bosons are considered.
Hence, the distribution of the probability amplitude is determined by the wave function $\Psi_{S}$
along with the amplitude of the spin projection probability.
The spinor structure of the wave function is reflected in the subindex $\Psi_{S}$,
where spin index $S$ is the short notation for the $N$ indexes of spin of each particle $S=\{s_{1}, ..., s_{N}\}$.
More accurately, the spin-1 particles are the vector particles,
where single particle wave function is the three component column,
while the spinor is the mathematical object invented for the spin-1/2 fermions.
The spin structure of the wave function is the direct product of $N$ three component columns
and the Hamiltonian $\hat{H}=\hat{H}_{SS'}$ acting as the $3N$-rank matrix.
The action of matrix $\hat{H}_{SS'}$ on the "spinor-vector" $\Psi_{S'}$ includes summation on subindex $S'$:
$\sum_{S'=s_{1}',...,s_{N}'}\hat{H}_{SS'}\Psi_{S'}$,
with $s_{i}=0$, $\pm1$,
but equation (\ref{BECfinTSpin1 Schrodinger equation micro}) does not include the summation
since the summation on the repeating index is assumed.

To specify the physical system under consideration requires the explicit form of the Hamiltonian of the system:
$$\hat{H}=\sum_{i=1}^{N}\biggl(\frac{\hat{\textbf{p}}^{2}_{i}}{2m}
+V_{ext}(\textbf{r}_{i},t)-p\hat{F}_{i,z}+q\hat{F}_{i,z}^{2}\biggr)$$
\begin{equation}\label{BECfinTSpin1 Hamiltonian micro}
+\frac{1}{2}\sum_{i,j\neq i}\biggl(U_{1}(\textbf{r}_{i}-\textbf{r}_{j})
+U_{2}(\textbf{r}_{i}-\textbf{r}_{j})\hat{\textbf{F}}_{i}\cdot\hat{\textbf{F}}_{j} \biggr),\end{equation}
where $m$ is the mass of atom,
$\hat{\textbf{p}}_{i}=-\imath\hbar\nabla_{i}$ is the momentum of i-th atom.
The first group of terms is the superposition of the singe-particle Hamiltonians,
which includes the linear Zeeman term proportional to $p$
and the quadratic Zeeman term proportional to $q$.
The last group of terms in the Hamiltonian (\ref{BECfinTSpin1 Hamiltonian micro})
contains two terms.
The scalar spinless part of the short-range boson-boson interaction $U_{1,ij}$
which equally contributes in interaction of atoms with same spin projection.
The spindependent part of the short-range boson-boson interaction $U_{2,ij}$ demonstrates
deviation from potential $U_{1,ij}$ for the interaction of atoms with different spin projections.
The explicit form of the short-range interaction is not specified.
No distinguish between bosons in the BEC state and bosons in other states is made at the microscopic level.
Separation of all bosons on BEC and normal fluid is made in terms of collective variables.

Hamiltonian (\ref{BECfinTSpin1 Hamiltonian micro}) contains the single-particle spin matrixes
$$\hat{\textbf{F}}_{i}=\{\hat{F}_{x}, \hat{F}_{y}, \hat{F}_{z}\}$$
\begin{equation}\label{BECfinTSpin1 spin matrixes} =\begin{array}{ccc}
                                                     \frac{1}{\sqrt{2}}\left(
                                                       \begin{array}{ccc}
                                                         0 & 1 & 0 \\
                                                         1 & 0 & 1 \\
                                                         0 & 1 & 0 \\
                                                       \end{array}
                                                     \right),
                                                      &
                                                      \frac{1}{\sqrt{2}}\left(
                                                       \begin{array}{ccc}
                                                         0 & -\imath & 0 \\
                                                         \imath & 0 & -\imath \\
                                                         0 & \imath & 0 \\
                                                       \end{array}
                                                     \right),
                                                     &
                                                     \left(
                                                       \begin{array}{ccc}
                                                         1 & 0 & 0 \\
                                                         0 & 0 & 0 \\
                                                         0 & 0 & -1 \\
                                                       \end{array}
                                                     \right).
                                                   \end{array}
\end{equation}

The microscopic distribution of spin-1 bosons in the physical space can be obtained by the projection of the distribution in the configurational space
\cite{Maksimov QHM 99}, \cite{Andreev PRA08}, \cite{Andreev 2001}, \cite{Andreev 1912}, \cite{Andreev LPL 18}, \cite{Andreev 2007}
\begin{equation}\label{BECfinTSpin1 concentration n definition}
n=\int \Psi_{S}^{\dagger}(R,t)\sum_{i=1}^{N}\delta(\textbf{r}-\textbf{r}_{i})\Psi_{S}(R,t)dR. \end{equation}
This definition contains the integration over the configurational space and the summation over the spin indexes.
Same operations are included in the definitions of other hydrodynamic functions presented below.

Each boson has probability to be in quantum state with different projections of spin.
Considering all bosons in the quasi-classic terms we can refer to number of bosons with different spin projections.
Therefore, the concentration of all bosons can be separated on the partial concentrations $n=n_{1}+n_{0}+n_{-1}$,
where subindexes $0$, $\pm1$ refer to the spin projections of bosons.

We consider system of bosons at the arbitrary temperature.
Moreover, we derive equations for the nonequilibrium systems,
where temperature changes in space and time.
Temperature is the characteristic of the collective behavior like the concentration
(\ref{BECfinTSpin1 concentration n definition}) and other hydrodynamic functions presented below.
Therefore, it is not presented in the many-particle wave function describing the microscopic dynamics of quantum system.

Let us calculate the temporal evolution of the concentration (\ref{BECfinTSpin1 concentration n definition}).
We consider the time derivative of the concentration.
The time derivative acts on the wave functions under integral
while the time derivatives of the wave function are replaced by the Hamiltonian
in accordance with the Schrodinger equation (\ref{BECfinTSpin1 Schrodinger equation micro}).
The straightforward calculations lead to the continuity equation
\begin{equation}\label{BECfinTSpin1 continuity equation via j}
\partial_{t}n+\nabla\cdot \textbf{j}=0. \end{equation}
The appearance of the continuity equation is rather obvious conclusion.
However, the derivation of the continuity equation provides the explicit definition for the current in terms of the many-particle wave function:
\begin{equation}\label{BECfinTSpin1 j def}
\textbf{j}=\frac{1}{2m}\int \sum_{i=1}^{N}\delta(\textbf{r}-\textbf{r}_{i})
[\Psi_{S}^{\dagger}(R,t)\hat{\textbf{p}}_{i}\Psi_{S}(R,t)+h.c.]dR, \end{equation}
where $h.c.$ stands for the Hermitian conjugation.
Vector $\textbf{j}$ is the probability current in the physical space
which reduces to the current of particles in the classical limit.
Moreover, vector $\textbf{j}$ is proportional to the momentum density of the quantum system.

Next step is the derivation of the momentum balance equation
which is the preliminary form of the Euler equation
\begin{equation}\label{BECfinTSpin1 j evol eq}
\partial_{t}j^{\alpha}+\partial_{\beta}\Pi^{\alpha\beta}=
-\frac{1}{m}n\partial^{\alpha}V_{ext}+\frac{1}{m}F_{int}^{\alpha}, \end{equation}
where
the summation on the repeating vector indexes is assumed,
and
$$\Pi^{\alpha\beta}=\frac{1}{4m^{2}}\int dR\sum_{i}\delta(\textbf{r}-\textbf{r}_{i})
[\Psi_{S}^{\dagger}(R,t)\hat{p}_{i}^{\alpha}\hat{p}_{i}^{\beta}\Psi_{S}(R,t)$$
\begin{equation} \label{BECfinTSpin1 Pi def}
+(\hat{p}_{i}^{\alpha}\Psi(R,t))_{S}^{\dagger}\hat{p}_{i}^{\beta}\Psi_{S}(R,t)+h.c.] \end{equation}
is the momentum flux,
and $F_{int}^{\alpha}$ is the force field describing the interaction between bosons.

The general form of the interparticle interaction force field appears in the following form:
$$F_{int}^{\alpha}=-\int (\partial^{\alpha}U_{1}(\textbf{r}-\textbf{r}'))n_{2}(\textbf{r},\textbf{r}',t)d\textbf{r}'$$
\begin{equation}\label{BECfinTSpin1 F via n2 S2}
-\int (\partial^{\alpha}U_{2}(\textbf{r}-\textbf{r}'))S_{2}^{\beta\beta}(\textbf{r},\textbf{r}',t)d\textbf{r}', \end{equation}
where
\begin{equation}\label{BECfinTSpin1 n 2 def} n_{2}(\textbf{r},\textbf{r}',t)=
\int \Psi_{S}^{\dagger}\sum_{i,j\neq i}\delta(\textbf{r}-\textbf{r}_{i})\delta(\textbf{r}'-\textbf{r}_{j})\Psi_{S}dR, \end{equation}
and
$$S_{2}^{\alpha\beta}(\textbf{r},\textbf{r}',t)=
\int \Psi_{S}^{\dagger}(R,t)\sum_{i,j\neq i}\delta(\textbf{r}-\textbf{r}_{i})\times$$
\begin{equation}\label{BECfinTSpin1 S 2 def} \times\delta(\textbf{r}'-\textbf{r}_{j})
[\hat{F}_{i}^{\alpha}\hat{F}_{j}^{\beta}\Psi(R,t)]_{S}dR, \end{equation}
with
$[\hat{\textbf{F}}_{i}\cdot\hat{\textbf{F}}_{j}\Psi(R,t)]_{S}$ stands for
$[\hat{\textbf{F}}_{s_{i}s_{i}'}\cdot\hat{\textbf{F}}_{s_{j}s_{j}'}\Psi_{s_{1},...,s_{i}',...,s_{j}',...,s_{N}}(R,t)$.

Approximate calculation of $F_{int}^{\alpha}$ for the short-range interaction in the first order by the interaction radius
\cite{Andreev PRA08}, \cite{Andreev IJMP B 13}, \cite{Andreev 2001}, \cite{Andreev LP 19}
leads to the following result
$$F_{int}^{\alpha}=-\frac{1}{2}g_{1}\partial^{\alpha}(2n_{n}^{2}+4n_{n}n_{B}+n_{B}^{2})$$
\begin{equation}\label{BECfinTSpin1 F via n n S S}
-\frac{1}{2}g_{2}\partial^{\alpha}(2\textbf{S}_{n}^{2}+4\textbf{S}_{n}\textbf{S}_{B}+\textbf{S}_{B}^{2}), \end{equation}
where
\begin{equation}\label{BECfinTSpin1 g i def} g_{i}=\int U_{i}(r)d\textbf{r} \end{equation}
are the interaction constants for the two parts of the short-range interaction potential
presented in the Hamiltonian (\ref{BECfinTSpin1 Hamiltonian micro}),
the subindex $B$ refers to the BEC, the subindex $n$ refers to the normal fluid,
and
\begin{equation}\label{BECfinTSpin1 S def} \textbf{S}(\textbf{r},t)=
\int \Psi_{S}^{\dagger}(R,t)\sum_{i}\delta(\textbf{r}-\textbf{r}_{i})
(\hat{\textbf{F}}_{i}\Psi(R,t))_{S}dR \end{equation}
is the spin density.
Moreover, expression (\ref{BECfinTSpin1 F via n n S S}) is presented in terms of two fluid model.
The functions describing the BEC and the normal fluid enter this equation in nonsymmetric form,
so they are not combined in the functions describing all bosons simultaneously.
This result for the force field gives additional motivation to split hydrodynamic equations on two subsystems.

Equation (\ref{BECfinTSpin1 F via n n S S}) has the following zero temperature limit
\begin{equation}\label{BECfinTSpin1 F via n n S S}
F_{int,BEC}^{\alpha}=-g_{1}n_{B}\partial^{\alpha}n_{B}
-g_{2}\textbf{S}_{B}\partial^{\alpha}\textbf{S}_{B}. \end{equation}

Presence of the spin density (\ref{BECfinTSpin1 S def}) leads to the necessity to derive the equation for its evolution
$$\partial_{t}S^{\alpha}+\partial_{\beta}J^{\alpha\beta}=
-\frac{p}{\hbar}\varepsilon^{\alpha z \gamma}S^{\gamma}
+2\frac{q}{\hbar} \varepsilon^{\alpha z \gamma}N^{z\gamma}$$
\begin{equation}\label{BECfinTSpin1 spin evolution}
+\frac{\varepsilon^{\alpha\beta\gamma}}{\hbar}
\int U_{2}(\textbf{r}-\textbf{r}')S_{2}^{\gamma\beta}(\textbf{r},\textbf{r}',t)d\textbf{r}',\end{equation}
where
\begin{equation}\label{BECfinTSpin1 J alpha beta def} J^{\alpha\beta}=\frac{1}{2m}\int \sum_{i=1}^{N}\delta(\textbf{r}-\textbf{r}_{i})
[\Psi_{S}^{\dagger}\hat{F}_{i}^{\alpha}\hat{p}_{i}^{\beta}\Psi_{S}+h.c.]dR \end{equation}
is the spin current giving the redistribution of particles with no influence of interaction,
and
\begin{equation}\label{BECfinTSpin1 nematic tensor def} N^{\alpha\beta}=\frac{1}{2}\int \sum_{i=1}^{N}\delta(\textbf{r}-\textbf{r}_{i})
\Psi_{S}^{\dagger}(\hat{F}_{i}^{\alpha}\hat{F}_{i}^{\beta}+\hat{F}_{i}^{\beta}\hat{F}_{i}^{\alpha})\Psi_{S}dR \end{equation}
is the density of the macroscopic nematic tensor.
It is defined as the quantum average of the nematic tensor operator
$\hat{N}_{i}^{\alpha\beta}=(1/2)(\hat{F}_{i}^{\alpha}\hat{F}_{i}^{\beta}+\hat{F}_{i}^{\beta}\hat{F}_{i}^{\alpha})$ of the $i$-th particle.
The nematic tensor is the independent function for the spin-1 bosons,
while for the spin-1/2 fermions it reduces to the concentration of particles.

The general form of the spin evolution equation (\ref{BECfinTSpin1 spin evolution}) contains the interparticle interaction in the last term.
However, its calculation in the first order by the interaction radius gives the zero value.
Hence, equation (\ref{BECfinTSpin1 spin evolution}) reduces to
\begin{equation}\label{BECfinTSpin1 spin evolution reduced}
\partial_{t}S^{\alpha}+\partial_{\beta}J^{\alpha\beta}=
-\frac{p}{\hbar}\varepsilon^{\alpha z \gamma}S^{\gamma}
+2\frac{q}{\hbar} \varepsilon^{\alpha z \gamma}N^{z\gamma}.\end{equation}
Therefore, no interaction gives contribution in the spin evolution equations.
The nontrivial evolution of spin density is related to the quadratic Zeeman effects presented via
novel hydrodynamic function: the nematic tensor $N^{\alpha\beta}$.
If we want to obtain the influence of interaction on the spin density evolution
we need to derive the nematic tensor evolution equation.
Hence, we consider the time derivative of function (\ref{BECfinTSpin1 nematic tensor def})
and use the Schrodinger equation (\ref{BECfinTSpin1 Schrodinger equation micro})
with Hamiltonian (\ref{BECfinTSpin1 Hamiltonian micro}) for the time derivatives of the microscopic many-particle wave function.
As the result we find the following equation:
$$\partial_{t}N^{\alpha\beta}+\partial_{\gamma}J_{N}^{\alpha\beta\gamma}
=\frac{p}{\hbar}\varepsilon^{z\alpha\gamma}N^{\beta\gamma}+\frac{p}{\hbar}\varepsilon^{z\beta\gamma}N^{\alpha\gamma}$$
$$-\frac{q}{2\hbar}(\varepsilon^{z\alpha\gamma}\delta^{\beta z}
+\varepsilon^{z\beta\gamma}\delta^{\alpha z})S^{\gamma}$$
$$+\frac{g_{2}}{\hbar}\biggl[ \varepsilon^{\beta\gamma\delta}
\biggl( 2N_{n}^{\alpha\delta}S_{n}^{\gamma} +2N_{n}^{\alpha\delta}S_{B}^{\gamma} +2N_{B}^{\alpha\delta}S_{n}^{\gamma}
+N_{B}^{\alpha\delta}S_{B}^{\gamma}\biggr)$$
\begin{equation}\label{BECfinTSpin1 N eq evol}
+\varepsilon^{\alpha\gamma\delta}
\biggl( 2N_{n}^{\beta\delta}S_{n}^{\gamma} +2N_{n}^{\beta\delta}S_{B}^{\gamma} +2N_{B}^{\beta\delta}S_{n}^{\gamma}
+N_{B}^{\beta\delta}S_{B}^{\gamma}\biggr)\biggr],
\end{equation}
where
$$J_{N}^{\alpha\beta\gamma}=\frac{1}{4m}\int \sum_{i=1}^{N}\delta(\textbf{r}-\textbf{r}_{i})
[\Psi_{S}^{\dagger}(R,t)\times$$
\begin{equation}\label{BECfinTSpin1 J a b g def}
\times(\hat{F}_{i}^{\alpha}\hat{F}_{i}^{\beta}+\hat{F}_{i}^{\beta}\hat{F}_{i}^{\alpha})\hat{p}_{i}^{\gamma}\Psi_{S}(R,t)+h.c.]dR \end{equation}
is the flux of the nematic tensor,
the subindex $N$ refers to the nematic tensor,
and
the last group of terms proportional to the second interaction constant
appears at the calculation of the following term in the first order by the interaction radius
$$\int dR\sum_{i,k\neq i}\delta(\textbf{r}-\textbf{r}_{i})U_{2,ik}\Psi_{S}^{\dagger}(R,t)\times$$
\begin{equation}\label{BECfinTSpin1 interaction for N}
\times(\varepsilon^{\beta\gamma\delta}\hat{N}_{i}^{\alpha\delta}\hat{F}_{k}^{\gamma}
+\varepsilon^{\alpha\gamma\delta}\hat{N}_{i}^{\beta\delta}\hat{F}_{k}^{\gamma})\Psi_{S}(R,t). \end{equation}

The general structure of hydrodynamic equations is obtained above.
However, we need to introduce the velocity field to give these equations more traditionally appearance.
Moreover, this allows to consider the structure of fluxes of all variables.

\subsection{Madelung transformation for the single-particle wave function}

Before we present the structure of the many-particle wave function
let us consider the single particle wave function.
For the spin-1 boson it is the three component column presenting the vector wave function
\begin{equation}\label{BECfinTSpin1 psi general vector} \psi(\textbf{r},t)=\left(
                                                          \begin{array}{c}
                                                            \psi_{+1} \\
                                                            \psi_{0} \\
                                                            \psi_{-1} \\
                                                          \end{array}
                                                        \right),
\end{equation}
where the amplitude $a(\textbf{r},t)$ and phase $S(\textbf{r},t)$ of the wave function can be introduced
together with the unit vector showing the spin state of the particle
\begin{equation}\label{BECfinTSpin1 psi via a S z} \psi=a e^{iS} \hat{z}(\Theta), \end{equation}
where we have unit vector $\hat{z}(\Theta)$ describing the possible spin states.

Making a step ahead we consider the example of the unit vector $\hat{z}(\Theta)$ applied for the description
of the ferromagnetic phase of spin-1 BECs
\begin{equation}\label{BECfinTSpin1 z via angles F} \hat{z}=e^{-\imath\chi}\left(
                                                              \begin{array}{c}
                                                                \cos^{2}(\theta/2) e^{-\imath\varphi} \\
                                                                \frac{1}{\sqrt{2}}\sin(\theta) \\
                                                                \sin^{2}(\theta/2) e^{\imath\varphi} \\
                                                              \end{array}
                                                            \right),
\end{equation}
with
$\Theta=\{\chi,\theta,\varphi\}$,
and
$\chi(\textbf{r},t)$, $\theta(\textbf{r},t)$, $\varphi(\textbf{r},t)$ are the scalar fields,
\emph{and}
$\hat{z}^{\dag}\hat{z}=1$.

Vector (\ref{BECfinTSpin1 z via angles F}) can be obtain by rotation of vector
\begin{equation}\label{BECfinTSpin1 z 100}\hat{z}=\left(
                                                              \begin{array}{c}
                                                                1 \\
                                                                0 \\
                                                                0 \\
                                                              \end{array}
                                                            \right),
\end{equation}
where the rotation is made by matrix
\begin{equation}\label{BECfinTSpin1 Rotation Matrix}
\left(
  \begin{array}{ccc}
    e^{-\imath\chi}\cos^{2}\frac{\theta}{2} e^{-\imath\varphi} &
    -\frac{e^{-\imath\varphi}}{\sqrt{2}} \sin\theta  & e^{\imath\chi}\cos^{2}\frac{\theta}{2} e^{-\imath\varphi} \\
    e^{-\imath\chi}\frac{1}{\sqrt{2}}\sin\theta & \cos\theta & e^{\imath\chi}\frac{1}{\sqrt{2}}\sin\theta \\
    e^{-\imath\chi}\sin^{2}\frac{\theta}{2} e^{\imath\varphi} &
    \frac{e^{\imath\varphi}}{\sqrt{2}} \sin\theta & e^{\imath\chi}\sin^{2}\frac{\theta}{2} e^{\imath\varphi} \\
  \end{array}
\right).
\end{equation}

The second phase of the spin-1 BECs is the polar phase.
It corresponds to the unit vector obtained by the rotation of vector
\begin{equation}\label{BECfinTSpin1 z 010} \hat{z}=\left(
                                                              \begin{array}{c}
                                                                0 \\
                                                                1 \\
                                                                0 \\
                                                              \end{array}
                                                            \right),
\end{equation}
which leads to
\begin{equation}\label{BECfinTSpin1 z P}\hat{z}=\left(
                                                              \begin{array}{c}
                                                                -\frac{1}{\sqrt{2}} \sin(\theta)e^{-\imath\varphi} \\
                                                                \cos(\theta) \\
                                                                \frac{1}{\sqrt{2}} \sin(\theta)e^{\imath\varphi} \\
                                                              \end{array}
                                                            \right),
\end{equation}


However, these are examples of unit vector $\hat{z}$ from the macroscopic physics.
Or it can be considered as the wave function for the single particle.
We need to make step back to the microscopic description.

\subsection{Introduction of the velocity field and the average spin field via the Madelung transformation}

For the many-particle wave function we have $3N$ dimensional configurational space.
Moreover, the spin part of the wave function is the direct product of $N$ columns of three component each (\ref{BECfinTSpin1 psi general vector}).
Here, we can express the amplitude $a(R,t)$ and phase $S(R,t)$
\begin{equation}\label{BECfinTSpin1 Psi a S zzz} \Psi_{S}(R,t)= a(R,t) e^{\imath S(R,t)}\bigotimes_{i=1}^{N}\hat{z}_{i} \end{equation}
where
$\bigotimes_{i=1}^{N}\hat{z}_{i}=\hat{z}_{1}\otimes \hat{z}_{2} \otimes... \otimes \hat{z}_{N}$,
and
$\hat{z}_{i}=\hat{z}_{i}(\Theta_{i})$,
with
$\Theta_{i}=\Theta_{i}(\textbf{r}_{i},t)$ is the set of three angular fields defined in three dimensional subspace of the configurational space.

Generalized Madelung decomposition (\ref{BECfinTSpin1 Psi a S zzz}) gives the following representation of the concentration:
\begin{equation}\label{BECfinTSpin1 n def via a} n=\int dR \sum_{i=1}^{N}\delta(\textbf{r}-\textbf{r}_{i})a^{2}(R,t). \end{equation}
It is independent of the explicit form of unit vectors $\hat{z}_{i}$.

Other hydrodynamic functions contain the contribution of unit vectors $\hat{z}_{i}$.
The generalized Madelung decomposition (\ref{BECfinTSpin1 Psi a S zzz}) provides the modified expression for the current
$$\textbf{j}=\frac{\hbar}{m}
\int dR \sum_{i=1}^{N}\delta(\textbf{r}-\textbf{r}_{i})a^{2}(R,t)\times$$
\begin{equation}\label{BECfinTSpin1 j def via a S}
\times(\partial_{i}^{\alpha}S(R,t)-\imath\hat{z}_{i}^{\dag}\partial_{i}^{\alpha}\hat{z}_{i}), \end{equation}
where the gradient of phase $S(R,t)$ existing for the spinless particles is shifted by the gradient of the unit vectors.

We have the following specification of equation (\ref{BECfinTSpin1 j def via a S}) for the ferromagnetic phase
$$\textbf{j}=\frac{\hbar}{m}
\int dR \sum_{i=1}^{N}\delta(\textbf{r}-\textbf{r}_{i})a^{2}(R,t)\times$$
\begin{equation}\label{BECfinTSpin1 j def via a S for F}
\times(\partial_{i}^{\alpha}S(R,t)-\partial_{i}^{\alpha}\chi_{i}-\cos\theta_{i}\partial_{i}^{\alpha}\varphi_{i}), \end{equation}
where the velocity of  $i$-th particle can be introduced in terms of components of unit vector $\hat{z}_{i}$ (\ref{BECfinTSpin1 z via angles F}):
$v_{i}^{\alpha}=\partial_{i}^{\alpha}S-\partial_{i}^{\alpha}\chi_{i}-\cos\theta_{i}\partial_{i}^{\alpha}\varphi_{i}$.
For the arbitrary unit vector $\hat{z}_{i}$ we also introduce the velocity of  $i$-th particle
$v_{i}^{\alpha}=\partial_{i}^{\alpha}S-\imath\hat{z}_{i}^{\dag}\partial_{i}^{\alpha}\hat{z}_{i}$.

Substitution of the wave function via the arbitrary unit vector $\hat{z}_{i}$ contains the contribution of spin of $i$-th particle
$\textbf{s}_{i}=\hat{z}_{i}^{\dag}\hat{\textbf{F}}_{i}\hat{z}_{i}$ under the integral on the configurational space
\begin{equation}\label{BECfinTSpin1 S def via a z} \textbf{S}=\int dR \sum_{i=1}^{N}\delta(\textbf{r}-\textbf{r}_{i})
(\hat{z}_{i}^{\dag}\hat{\textbf{F}}_{i}\hat{z}_{i}) a^{2}(R,t). \end{equation}
Other unit vectors $\hat{z}_{j\neq i}$ exclude themselves via the normalization.

At the full polarization of the spin of each particle $\textbf{s}_{i}$ appear via the unit vector:
\begin{equation}\label{BECfinTSpin1 S def via a z F} \textbf{S}=\int dR \sum_{i=1}^{N}\delta(\textbf{r}-\textbf{r}_{i})\textbf{n}_{i}a^{2}(R,t), \end{equation}
where
$\textbf{s}_{i}=\textbf{n}_{i}=\textbf{n}(\textbf{r}_{i},t)$$=\{\cos\varphi_{i}\sin\theta_{i}, \sin\varphi_{i}\sin\theta_{i} ,\cos\theta_{i}\}$
is the unit vector
which shows the contribution of the spin of $i$-th particle in the spin density vector field.
We have unit module of each spin,
but spin can have different directions relatively each other.
Hence, it can correspond to the partial spin polarization.
Moreover, the partial spin polarization of the system can be caused by the partial spin polarization of each particle.
The ferromagnetic phase exists at the full polarization of the spin of each particle $\textbf{s}_{i}$
which have same spin polarization.
In the polar phase $\textbf{s}_{i}=0$.
Hence, $\textbf{S}_{p}=0$,
where subindex $p$ stands for the indication of the polar phase.

The last in the line of major hydrodynamic functions necessary for the description of the spin-1 BECs
is the density of the nematic tensor $N^{\alpha\beta}$.
It can be represented by the single-particle nematic tensor
$n^{\alpha\beta}_{i}=$$(1/2)\hat{z}_{i}^{\dag}(\hat{F}_{i}^{\alpha}\hat{F}_{i}^{\beta}$$+\hat{F}_{i}^{\beta}\hat{F}_{i}^{\alpha})\hat{z}_{i}$
under the integral on the configurational space
\begin{equation}\label{BECfinTSpin1 N def via a z} N^{\alpha\beta}=\frac{1}{2}\int dR \sum_{i=1}^{N}\delta(\textbf{r}-\textbf{r}_{i})
(\hat{z}_{i}^{\dag}(\hat{F}_{i}^{\alpha}\hat{F}_{i}^{\beta}+\hat{F}_{i}^{\beta}\hat{F}_{i}^{\alpha})\hat{z}_{i}) a^{2}(R,t). \end{equation}
Let us consider the structure of the single-particle nematic tensor $n^{\alpha\beta}_{i}$.
At the full polarization of the spin of each particle $\textbf{s}_{i}=\textbf{n}_{i}$
we have the following simplification of the single-particle nematic tensor
$n^{\alpha\beta}_{i}=(\delta^{\alpha\beta}+n_{i}^{\alpha}n_{i}^{\beta})/2$.
Or it can be rewritten via the spin of $i$-th particle
$n^{\alpha\beta}_{i}=(\delta^{\alpha\beta}+s_{i}^{\alpha}s_{i}^{\beta})/2$.
The transition to the nematic tensor density requires to consider the quantum average of this function on the many-particle wave function.
Hence, in general, the second term does not give the product of two spin densities $\textbf{S}$.
Corresponding nematic tensor is discussed in this section below.


\subsection{Transformation of the fluxes at the Madelung transformation}

Above we consider the time evolution of the concentration $n$, the current $\textbf{j}$,
the spin density $\textbf{S}$, and the nematic tensor $N^{\alpha\beta}$.
The interaction influence their evolution.
However, the fluxes of these functions give their redistribution even at the zero interaction.
To understand the structure of the quantum hydrodynamic equations
we need to consider the representation of the fluxes of basic hydrodynamic variables
at the generalized Madelung decomposition (\ref{BECfinTSpin1 Psi a S zzz}).
The current $\textbf{j}$ is the flux of concentration $n$.
Hence, one flux is considered.
We need to consider the momentum flux $\Pi^{\alpha\beta}$ (\ref{BECfinTSpin1 Pi def}),
the spin current $J^{\alpha\beta}$ (\ref{BECfinTSpin1 J alpha beta def}),
and the flux of nematic tensor $J_{N}^{\alpha\beta\gamma}$ (\ref{BECfinTSpin1 J a b g def}).

We start this part of analysis with the representation of the spin current $J^{\alpha\beta}$ (\ref{BECfinTSpin1 J alpha beta def}):
$$J^{\alpha\beta}=\frac{\hbar}{m}\int dR \sum_{i=1}^{N}\delta(\textbf{r}-\textbf{r}_{i})s_{i}^{\alpha}(\partial_{i}^{\beta}S(R,t))a^{2}(R,t)$$
\begin{equation}\label{BECfinTSpin1 J alpha beta def via a z I}
+\frac{1}{2m}\int dR \sum_{i=1}^{N}\delta(\textbf{r}-\textbf{r}_{i})a^{2}(R,t)(z_{i}^{\dag}\hat{F}^{\alpha}_{i}\hat{p}_{i}^{\beta}z_{i}+h.c.).
\end{equation}
The first term in the spin current contains the spin of $i$-th particle,
but it contains a part of the velocity of $i$-th particle.
Let us represent $\partial_{i}^{\beta}S(R,t)$ via the velocity $v_{i}^{\alpha}$.
Hence, we have term which has clear physical meaning of the spin current
via the product of spin $s_{i}^{\alpha}$ and velocity of the same particle $v_{i}^{\beta}$:
$$J^{\alpha\beta}=\int dR \sum_{i=1}^{N}\delta(\textbf{r}-\textbf{r}_{i})s_{i}^{\alpha}v_{i}^{\beta}a^{2}(R,t)$$
$$+\frac{\imath\hbar}{m}\int dR \sum_{i=1}^{N}\delta(\textbf{r}-\textbf{r}_{i})s_{i}^{\alpha}(z_{i}^{\dag}\partial_{i}^{\beta}z_{i})a^{2}(R,t)$$
\begin{equation}\label{BECfinTSpin1 J alpha beta def via a z II}
+\frac{1}{2m}\int dR \sum_{i=1}^{N}\delta(\textbf{r}-\textbf{r}_{i})a^{2}
(R,t)(z_{i}^{\dag}\hat{F}^{\alpha}_{i}\hat{p}_{i}^{\beta}z_{i}+h.c.).
\end{equation}
Hence, the first term can be called the quasi-classic part of the spin-current $J^{\alpha\beta}_{qc}$,
while other terms are the quantum corrections.


The second example of the fluxes is the momentum flux $\Pi^{\alpha\beta}$ (\ref{BECfinTSpin1 Pi def}).
Let us present the result of application of Madelung decomposition (\ref{BECfinTSpin1 Psi a S zzz}):
$$\Pi^{\alpha\beta}=\int dR \sum_{i=1}^{N}\delta(\textbf{r}-\textbf{r}_{i})a^{2}(R,t)v_{i}^{\alpha}v_{i}^{\beta}$$
$$+\frac{\hbar^{2}}{2m^{2}}\int dR \sum_{i=1}^{N}\delta(\textbf{r}-\textbf{r}_{i})
(\partial^{\alpha}_{i}a\cdot\partial^{\beta}_{i}a-a\partial^{\alpha}_{i}\partial^{\beta}_{i}a)$$
$$+\frac{\hbar^{2}}{4m^{2}}\int dR \sum_{i=1}^{N}\delta(\textbf{r}-\textbf{r}_{i})a^{2}
(\partial^{\alpha}_{i}z_{i}^{\dag}\cdot\partial^{\beta}_{i}z_{i}$$
\begin{equation}\label{BECfinTSpin1 Pi def via a v z}
+\partial^{\beta}_{i}z_{i}^{\dag}\cdot\partial^{\alpha}_{i}z_{i}
+4z_{i}^{\dag}\partial^{\alpha}_{i}z_{i}\cdot z_{i}^{\dag}\partial^{\beta}_{i}z_{i}). \end{equation}
The first term presents the quasi-classic part of the momentum flux $\Pi_{qc}^{\alpha\beta}$
which exists in the classic limit(if $\hbar\rightarrow0$).
Other terms are the quantum corrections.
The second term on the right-hand side of equation (\ref{BECfinTSpin1 Pi def via a v z})
is the spinless quantum correction.
The last group of terms is the spin related quantum correction.


Last example of the fluxes is the flux of nematic tensor:
$$J_{N}^{\alpha\beta\gamma}
=\int dR \sum_{i=1}^{N}\delta(\textbf{r}-\textbf{r}_{i})a^{2}(R,t)n_{i}^{\alpha\beta}v_{i}^{\gamma}$$
$$+\frac{\imath\hbar}{m}\int dR \sum_{i=1}^{N}\delta(\textbf{r}-\textbf{r}_{i})
n_{i}^{\alpha\beta}(z_{i}^{\dag}\partial_{i}^{\gamma}z_{i})a^{2}(R,t)$$
\begin{equation}\label{BECfinTSpin1 J a b g def via a z}
+\frac{1}{2m}\int dR \sum_{i=1}^{N}\delta(\textbf{r}-\textbf{r}_{i})
a^{2}(R,t)(z_{i}^{\dag}\hat{N}^{\alpha\beta}_{i}\hat{p}_{i}^{\gamma}z_{i}+h.c.).
\end{equation}
It also has the quasi-classical and quantum parts.


The "velocities" of different quantum particles give different contributions in the combined current $\textbf{j}$ (\ref{BECfinTSpin1 j def via a S}).
The current $\textbf{j}$ allows to introduce the average velocity (the velocity field) $\textbf{v}=\textbf{j}/n$,
which is the vector field showing local average velocity.
The deviation of velocity of each particle from the average velocity $\textbf{u}_{i}=\textbf{v}_{i}-\textbf{v}$ is introduced.
Chaotic variations from the average velocity can be associated with the thermal motion,while the temperature itself is the average square of velocity of the chaotic part of motion.
Hence, the temperature is proportional to the average of $\textbf{u}_{i}^{2}$.

Similar steps can be made for the analysis of structure of the spin density (\ref{BECfinTSpin1 S def via a z}).
The average spin can be introduced $\textbf{s}=\textbf{S}/n$.
Therefore, the deviation can be introduced $\textbf{t}_{i}=\textbf{s}_{i}-\textbf{s}$ as well.
Moreover, same separation can be made for the nematic tensor $N^{\alpha\beta}$:
$n_{i}^{\alpha\beta}=N^{\alpha\beta}/n+\nu_{i}^{\alpha\beta}$,
where $\nu_{i}^{\alpha\beta}$ is the deviation of the nematic tensor of $i$-th particle from the average value $N^{\alpha\beta}/n$.
The definitions of the spin density (\ref{BECfinTSpin1 S def}) and the nematic tensor density (\ref{BECfinTSpin1 nematic tensor def})
show that the average values of the deviations $\textbf{t}_{i}$ and $\nu_{i}^{\alpha\beta}$ are equal to zero.

Let us consider the structure of the nematic tensor density in the ferromagnetic phase.
Above we find the expression for the nematic tensor of $i$-th particle
via the spin of $i$-th particle in the ferromagnetic phase of the spin-1 bosons
$n^{\alpha\beta}_{i}=(\delta^{\alpha\beta}+s_{i}^{\alpha}s_{i}^{\beta})/2$.
We substitute the spin via the average spin $\textbf{s}=\textbf{S}/n$ and the deviation from the average
$\textbf{t}_{i}=\textbf{s}_{i}-\textbf{S}/n$.
Hence, the nematic tensor of $i$-th particle transforms to
$n^{\alpha\beta}_{i}=(1/2)(\delta^{\alpha\beta}+S^{\alpha}S^{\beta}/n^{2}$
$+(S^{\alpha}/n)t_{i}^{\beta}+(S^{\beta}/n)t_{i}^{\alpha} +t_{i}^{\alpha}t_{i}^{\beta})$.
The last term is constructed of the parameters defined in the local frame.
So, it is associated with the $\nu_{i}^{\alpha\beta}$,
which is the deviation of the nematic tensor of $i$-th particle from the average value $N^{\alpha\beta}/n$
We use it to find the nematic tensor density
$N^{\alpha\beta}=\int dR \sum_{i}\delta(\textbf{r}-\textbf{r}_{i})n^{\alpha\beta}_{i}a^{2}$.
Functions $S^{\alpha}$, $n$ can be easily extracted from under the integral.
We also include
$\int dR \sum_{i}\delta(\textbf{r}-\textbf{r}_{i})t^{\alpha}_{i}a^{2}=0$.
Hence, we find
$N^{\alpha\beta}=(1/2)n(\delta^{\alpha\beta}+S^{\alpha}S^{\beta}/n^{2})$
$+(1/2)\int dR \sum_{i}\delta(\textbf{r}-\textbf{r}_{i})\nu^{\alpha\beta}_{i}a^{2}$.
It is considered above that the module of the spin of each particle $s^{\alpha}_{i}$ is equal to one in the ferromagnetic phase.
Moreover, if we need to get the ferromagnetic state constructed of number of spins all of them should be parallel to each other.
Hence, there is no deviation of each spin from the average $t^{\alpha}_{i}=0$.
Therefore, the last term in the expression for the nematic tensor density $N^{\alpha\beta}$ is equal to zero.
Finally we obtain the well-known expression
$N^{\alpha\beta}=(1/2)(n\delta^{\alpha\beta}+S^{\alpha}S^{\beta}/n)$
(see for instance \cite{Yukawa PRA 12} in text after equation 31).

The first term in equation (\ref{BECfinTSpin1 Pi def via a v z}) is the quasi-classic part of the momentum flux.
We substitute the velocity via two terms $\textbf{v}_{i}=\textbf{u}_{i}+\textbf{v}$.
Next, we use that the average of the velocity in the comoving frame $\textbf{u}_{i}$ is equal to zero.
Therefore, the quasi-classic part of the momentum flux appears as the sum of two terms:
\begin{equation}\label{BECfinTSpin1 Pi via n v p}
\Pi_{qc}^{\alpha\beta}=nv^{\alpha}v^{\beta}+p^{\alpha\beta}, \end{equation}
where $p^{\alpha\beta}$ is the pressure with the following definition
\begin{equation}\label{BECfinTSpin1 pressure def}
p^{\alpha\beta}=\int dR \sum_{i=1}^{N}\delta(\textbf{r}-\textbf{r}_{i})a^{2}(R,t)u_{i}^{\alpha}u_{i}^{\beta}. \end{equation}
The nonequilibrium temperature scalar field is the average of the square of the velocity of the chaotic motion.
Hence, equation (\ref{BECfinTSpin1 pressure def}) shows that
the temperature is proportional to the trace of pressure $p^{\alpha\beta}$:
temperature is $T=p^{\beta\beta}/3n$,
where
$p^{\beta\beta}=p^{xx}+p^{yy}+p^{zz}$,
and
$p^{\beta\beta}=3p$
for the equal equal diagonal elements presented by $p$.

Same separation appears for the quasi-classic parts of the spin current and the nematic tensor flux
\begin{equation}\label{BECfinTSpin1 J alpha beta via S v} J_{qc}^{\alpha\beta}=S^{\alpha}v^{\beta}+J_{th}^{\alpha\beta}, \end{equation}
where we have the thermal part of the spin current tensor
\begin{equation}\label{BECfinTSpin1 J th alpha beta} J_{th}^{\alpha\beta}=
\int dR \sum_{i=1}^{N}\delta(\textbf{r}-\textbf{r}_{i})a^{2}(R,t)t_{i}^{\alpha}u_{i}^{\beta}, \end{equation}
\emph{and}
\begin{equation}\label{BECfinTSpin1 J alpha beta gamma via N v}
J_{N,qc}^{\alpha\beta\gamma}=N^{\alpha\beta}v^{\gamma}+J_{N,th}^{\alpha\beta\gamma}, \end{equation}
where we have the thermal part of the nematic tensor flux
\begin{equation}\label{BECfinTSpin1 J th alpha beta gamma} J_{N,th}^{\alpha\beta\gamma}=
\int dR \sum_{i=1}^{N}\delta(\textbf{r}-\textbf{r}_{i})a^{2}(R,t)\nu_{i}^{\alpha\beta}u_{i}^{\gamma}. \end{equation}

Tensors $J_{th}^{\alpha\beta}$ and $J_{N,th}^{\alpha\beta\gamma}$ similarly to the pressure $p^{\alpha\beta}$
are related to the thermal motion of the quantum particles.
However, the thermal spin current and the thermal nematic tensor flux includes the chaotic motion of spin orientation.

Below, we present the set of two non-linear Schrodinger equations for the BEC and the normal fluid.
It presents the minimal coupling model similar
to the set of non-linear Schrodinger equations existing in literature for the spin-0 bosons \cite{Dalfovo RMP 99}.
So, the contribution of the pressure of the normal fluid is neglected as well.

Equations (\ref{BECfinTSpin1 Pi via n v p}), (\ref{BECfinTSpin1 J alpha beta via S v}) and (\ref{BECfinTSpin1 J alpha beta gamma via N v})
presents the flux of momentum, the flux of spin, and the flux of the nematic tensor in the quasi-classic limit,
so the quantum terms are neglected there.
These expressions are obtained for the arbitrary temperatures.
However, we are interested in the small temperature limit.
First, consider the zero temperature limit.
Hence, no deviations of the velocity, spin, and the nematic tensor from the local average exist.
In our notations the zero temperature leads to
$u_{i}^{\alpha}=0$, $t_{i}^{\alpha}=0$, and $\nu_{i}^{\alpha\beta}=0$.
It also leads to $p_{B}^{\alpha\beta}=0$, $J_{th,B}^{\alpha\beta}=0$, and
$J_{th,B}^{\alpha\beta\gamma}=0$,
where the subindex $B$ refers to the BEC state.
In the nonzero temperature regime in the small temperature limit
functions
$p^{\alpha\beta}$, $J_{th}^{\alpha\beta}$, and
$J_{th}^{\alpha\beta\gamma}$ have nonzero value.
Moreover, they are associated with the normal fluid.
The value of parameters
$u_{i}^{\alpha}$, $t_{i}^{\alpha}$, and $\nu_{i}^{\alpha\beta}$
have the following behavior.
If we consider their value from the quasi-classic point of view we can state that for some particles being in the BEC state these parameters are equal to zero.
While for other particles being in the excited states these parameters are non zero,
and it leads to nonzero value of functions $p^{\alpha\beta}$, $J_{th}^{\alpha\beta}$, and
$J_{th}^{\alpha\beta\gamma}$.
However, more accurate quantum picture is a bit different.
Each boson at small temperature has probability to be in the BEC state and in some excited states.
Hence, parameters $u_{i}^{\alpha}$, $t_{i}^{\alpha}$, and $\nu_{i}^{\alpha\beta}$ can be nonzero for all particles,
but their values are reduced by the partial probability to find each particle in the excited states.





\section{Transition to the two-fluid hydrodynamics}

Basic hydrodynamic functions are additive on the particles on the system.
Therefore, they are additive on the subsystems.
So, we have the separation of the concentration $n=n_{B}+n_{n}$,
where subindex $B$ refers to the Bose-Einstein condensate,
and subindex $n$ refers to the normal fluid.
Same is true for the current $\textbf{j}=\textbf{j}_{B}+\textbf{j}_{n}$,
the momentum flux $\Pi^{\alpha\beta}=\Pi^{\alpha\beta}_{B}+\Pi^{\alpha\beta}_{n}$,
the spin density $\textbf{S}=\textbf{S}_{B}+\textbf{S}_{n}$,
the spin current $J^{\alpha\beta}=J^{\alpha\beta}_{B}+J^{\alpha\beta}_{n}$,
the nematic tensor $N^{\alpha\beta}=N^{\alpha\beta}_{B}+N^{\alpha\beta}_{n}$,
and the flux of the nematic tensor
$J_{N}^{\alpha\beta\gamma}=J^{\alpha\beta\gamma}_{N,B}+J^{\alpha\beta\gamma}_{N,n}$.
Logically, first we separate the basic hydrodynamic functions on two subsystems.
Second, we introduce the velocity field in each subsystem in accordance with the method described above.

It leads to the following set of equations
\begin{equation}\label{BECfinTSpin1 continuity equation via j a}
\partial_{t}n_{a}+\nabla\cdot \textbf{j}_{a}=0, \end{equation}
\begin{equation}\label{BECfinTSpin1 j evol eq a}
\partial_{t}j_{a}^{\alpha}+\partial_{\beta}\Pi_{a}^{\alpha\beta}=
-\frac{1}{m}n_{a}\partial^{\alpha}V_{ext}+\frac{1}{m}F_{int,a}^{\alpha}, \end{equation}
\begin{equation}\label{BECfinTSpin1 spin evolution reduced a}
\partial_{t}S_{a}^{\alpha}+\partial_{\beta}J_{a}^{\alpha\beta}=
-\frac{p}{\hbar}\varepsilon^{\alpha z \gamma}S_{a}^{\gamma}
+\frac{2q}{\hbar} \varepsilon^{\alpha z \gamma}N_{a}^{z\gamma},\end{equation}
and
$$\partial_{t}N_{a}^{\alpha\beta}+\partial_{\gamma}J_{N,a}^{\alpha\beta\gamma}
=\frac{p}{\hbar}\varepsilon^{z\alpha\gamma}N_{a}^{\beta\gamma}+\frac{p}{\hbar}\varepsilon^{z\beta\gamma}N_{a}^{\alpha\gamma}$$
\begin{equation}\label{BECfinTSpin1 N eq evol a}
-\frac{q}{2\hbar}(\varepsilon^{z\alpha\gamma}\delta^{\beta z}
+\varepsilon^{z\beta\gamma}\delta^{\alpha z})S_{a}^{\gamma}+F_{N,a}^{\alpha\beta},
\end{equation}
where $a$ stands for $B$ and $n$,
$F_{int,a}^{\alpha}$ is the force field acting on species $a$ being caused by both species,
$F_{N,a}^{\alpha\beta}$ is the second rank force field tensor giving the contribution of the short-range interaction in the nematic tensor evolution equation
which acts on species $a$.

Partial fluxes can be represented via the partial velocity fields in accordance with equations derived above:
$\textbf{j}=\textbf{j}_{B}+\textbf{j}_{n}=n_{B}\textbf{v}_{B}+n_{n}\textbf{v}_{n}$,
$\Pi^{\alpha\beta}_{a}=n_{a}v_{a}^{\alpha}v_{a}^{\beta}+p_{a}^{\alpha\beta}$,
$p_{B}^{\alpha\beta}=0$,
$J^{\alpha\beta}_{a}=S_{a}^{\alpha}v_{a}^{\beta}$,
$J^{\alpha\beta\gamma}_{N,a}=N_{a}^{\alpha\beta}v_{a}^{\gamma}$,
where the thermal part of the spin current,
and the nematic tensor flux are neglected.

The force fields existing in the partial Euler equations
appear from the expression (\ref{BECfinTSpin1 F via n n S S}),
where the function under the derivative appears as the source of the force,
while the second function in the term present the subsystem moving under action of this force
$$F_{int,B}^{\alpha}=
-g_{1}n_{B}\partial^{\alpha}n_{B}
-2g_{1}n_{B}\partial^{\alpha}n_{n}$$
\begin{equation}\label{BECfinTSpin1}
-g_{2}\textbf{S}_{B}\partial^{\alpha}\textbf{S}_{B}
-2g_{2}\textbf{S}_{B}\partial^{\alpha}\textbf{S}_{n},
\end{equation}
and
$$F_{int,n}^{\alpha}=-2g_{1}n_{n}\partial^{\alpha}n_{B}
-2g_{1}n_{n}\partial^{\alpha}n_{n}$$
\begin{equation}\label{BECfinTSpin1}
-2g_{2}\textbf{S}_{n}\partial^{\alpha}\textbf{S}_{B}
-2g_{2}\textbf{S}_{n}\partial^{\alpha}\textbf{S}_{n}. \end{equation}

The second rank force field tensors have the following form in accordance with the separation of equation (\ref{BECfinTSpin1 N eq evol}):
$$F_{N,B}^{\alpha\beta}=\frac{g_{2}}{\hbar}\biggl[ \varepsilon^{\beta\gamma\delta}
N_{B}^{\alpha\delta}\biggl( 2S_{n}^{\gamma}+S_{B}^{\gamma} \biggr)$$
\begin{equation}\label{BECfinTSpin1 F a b BEC} +\varepsilon^{\alpha\gamma\delta}
N_{B}^{\beta\delta}\biggl( 2S_{n}^{\gamma}+S_{B}^{\gamma}\biggr)\biggr], \end{equation}
and
$$F_{N,n}^{\alpha\beta}=2\frac{g_{2}}{\hbar}\biggl[ \varepsilon^{\beta\gamma\delta}
N_{n}^{\alpha\delta}\biggl(S_{n}^{\gamma}+S_{B}^{\gamma} \biggr)$$
\begin{equation}\label{BECfinTSpin1 F a b NF} +\varepsilon^{\alpha\gamma\delta}
N_{n}^{\beta\delta}\biggl(S_{n}^{\gamma} +S_{B}^{\gamma} \biggr)\biggr]. \end{equation}
In equations
(\ref{BECfinTSpin1 F a b BEC}) and (\ref{BECfinTSpin1 F a b NF})
the spin densities $S_{B}$ and $S_{n}$
present the source of the force field acting on the nematic tensor of BEC in
(\ref{BECfinTSpin1 F a b BEC})
and the normal fluid in (\ref{BECfinTSpin1 F a b NF}).



Further calculation of the quantum terms in
(\ref{BECfinTSpin1 J alpha beta def via a z II}),
(\ref{BECfinTSpin1 Pi def via a v z}),
(\ref{BECfinTSpin1 J a b g def via a z})
allows to obtain their approximate explicit forms in term of the hydrodynamic functions.
Corresponding results for the BEC can be found in Ref. \cite{Yukawa PRA 12}
(see eq. 24 for the spin current,
eq. 26 for the current of the nematic tensor,
the divergence of the momentum flux is presented in the Euler equation (28),
its quantum part is proportional to the square of the Planck constant).
Same structures can be approximately used for the normal fluid as the equations of state for the corresponding functions.
It is necessary to mention that the quantum terms are required
to get NLSEs we need to use the explicit form of the quantum terms.


\subsection{Nonlinear Schrodinger equations}

The Gross-Pitaevskii equation is the famous example of the non-linear Schrodinger equations
used for the modeling of spin-0 neutral BECs.

If we consider the microscopic dynamics of spin-0 BECs
we can derive the quantum hydrodynamic equations by the method presented above (see also \cite{Andreev PRA08}).
Further construction of the macroscopic wave function
using the hydrodynamic functions allows to derive the Gross-Pitaevskii equation from the quantum hydrodynamic equations.
This step complites the derivation of the Gross-Pitaevskii equation from the microscopic quantum motion by the quantum hydrodynamic method
\cite{Maksimov QHM 99}, \cite{MaksimovTMP 2001}, \cite{Andreev PRA08}.

If we consider particles with the nonzero spin
we do not derive the non-linear Schrodinger equation from the single fluid (if we consider the BEC only) model.
However, we construct the non-linear Schrodinger equation
in order to give same hydrodynamic equations like the quantum hydrodynamics obtained from the microscopic motion.
Same method is used to obtain the pair of the non-linear Schrodinger equations for the BEC and the normal fluid of the spin-1 ultracold bosons.

Therefore, the non-linear Schrodinger equations have the following form:
$$\imath\hbar\partial_{t}\hat{\Phi}_{B}=\Biggl(-\frac{\hbar^{2}\nabla^{2}}{2m}+V_{ext}
-p\hat{F}_{z}+q\hat{N}_{zz}$$
\begin{equation}\label{BECfinTSpin1 GP B}
+g(n_{B}+2n_{n})+g_{2}(\textbf{S}_{B}+2\textbf{S}_{n})\hat{\textbf{F}}\Biggr)\hat{\Phi}_{B}
,\end{equation}
and
$$\imath\hbar\partial_{t}\hat{\Phi}_{n}=\Biggl(-\frac{\hbar^{2}\nabla^{2}}{2m}+V_{ext}
-p\hat{F}_{z}+q\hat{N}_{zz}$$
\begin{equation}\label{BECfinTSpin1 GP like nf}
+2g(n_{B}+n_{n})+2g_{2}(\textbf{S}_{B}+\textbf{S}_{n})\hat{\textbf{F}}\Biggr)\hat{\Phi}_{n},\end{equation}
where
the three component vector macroscopic wave functions have the following structure
$\hat{\Phi}_{a}=\sqrt{n_{a}}e^{\imath m\phi_{a}/\hbar}\hat{z}_{a}$.
The concentrations $n_{a}$ is the square of module of the macroscopic wave function
$n_{a}=\hat{\Phi}_{a}^{\dag}\hat{\Phi}_{a}$.
The spin densities $\textbf{S}_{a}$ are also represented via the macroscopic wave function
$\textbf{S}_{a}=\hat{\Phi}_{a}^{\dag}\hat{\textbf{F}}\hat{\Phi}_{a}$.
Other hydrodynamic functions like $j^{\alpha}$, $\Pi^{\alpha\beta}$, $N^{\alpha\beta}$, $J^{\alpha\beta}$, and $J_{N}^{\alpha\beta\gamma}$
are also expressed via the macroscopic wave functions $\hat{\Phi}_{a}$.
Corresponding equations are shown in Appendix.

\section{Collective excitations for the ferromagnetic phase: quasi-classic regime}

We focused on the part of the spectrum of the bulk collective excitations caused by the interparticle interaction.
Hence, the spin textures related to the quantum kinematic motion is neglected.

The ferromagnetic equilibrium state (phase) is considered.
Hence, the nematic tensor density is constructed of the concentration of bosons $n_{a}$ and the spin density $\textbf{S}_{a}$.

\subsection{Zero temperature limit}

Before we present the spectrum of the collective excitations existing at the finite temperatures
we calculate the zero temperature spectrum appearing from the presented model of spin-1 BECs.

Let us consider the equilibrium state of uniform $n_{B0}=const$
and the macroscopically motionless $\textbf{v}_{B0}=0$ BECs.
In the ferromagnetic phase the spin density is also nonzero $\textbf{S}_{B0}\neq0$.

In order to study the small amplitude collective excitations
we present the approximate form of the hydrodynamic equations containing terms up to first order on the small amplitude:

\begin{equation}\label{BECfinTSpin1 cont lin} \partial_{t}\delta n_{B}+n_{0B}\nabla\delta \textbf{v}_{B}=0, \end{equation}
\begin{equation}\label{BECfinTSpin1 Euler lin} mn_{0B}\partial_{t}\delta \textbf{v}_{B}
=-g_{1}n_{0B}\nabla n_{B}
-g_{2}S_{0B}^{\beta}\nabla\delta S_{B}^{\gamma}, \end{equation}
and
$$\partial_{t}\delta S_{B}^{\alpha}+S_{0B}^{\alpha}(\nabla\delta \textbf{v}_{B})$$
\begin{equation}\label{BECfinTSpin1 Spin evol lin}
=-\frac{p}{\hbar}\varepsilon^{\alpha z\gamma}(S^{\gamma}_{0B}+\delta S_{B}^{\gamma})
+2\frac{q}{\hbar}\varepsilon^{\alpha z\gamma}(N_{0B}^{z\gamma}+\delta N_{B}^{z\gamma}), \end{equation}
where
\begin{equation}\label{BECfinTSpin1 N a b via n S} N_{B}^{\alpha\beta}=\frac{1}{2}\delta^{\alpha\beta}n_{B}+\frac{1}{2}S_{B}^{\alpha}S_{B}^{\beta}/n_{B}. \end{equation}
It gives the equilibrium expression
\begin{equation}\label{BECfinTSpin1 N eq a b via n S}
N_{0B}^{\alpha\beta}=\frac{1}{2}\delta^{\alpha\beta}n_{0B}+\frac{1}{2}S_{0B}^{\alpha}S_{0B}^{\beta}/n_{0B} ,\end{equation}
and the linear perturbation
$$\delta N_{B}^{\alpha\beta}=\frac{1}{2}\delta^{\alpha\beta}\delta n_{B}
+\frac{1}{2}S_{0B}^{\alpha}\delta S_{B}^{\beta}/n_{0B} $$
\begin{equation}\label{BECfinTSpin1 N pert a b via n S}
+\frac{1}{2}S_{0B}^{\beta}\delta S_{B}^{\alpha}/n_{0B}
-\frac{1}{2}(S_{0B}^{\alpha}S_{0B}^{\beta}/n_{0B}^{2})\delta n_{B}. \end{equation}

Equilibrium part of equation (\ref{BECfinTSpin1 Spin evol lin}) existing on the right-hand side and has the following form
\begin{equation}\label{BECfinTSpin1} \varepsilon^{\alpha z\gamma}(-p S^{\gamma}_{0B}+q\delta^{z\gamma}+qS_{0B}^{z} S_{0B}^{\gamma}/n_{0B}). \end{equation}
The second term is equal to zero since $\varepsilon^{\alpha z\gamma}\delta^{z\gamma}= \varepsilon^{\alpha zz}=0$.
The rest should be equal to zero in order to satisfy the hydrodynamic equations with the chosen equation of state.
It is satisfied for $\textbf{S}_{0B}=S_{0B} \textbf{e}_{z}$.
Moreover, all spin are parallel to each other in the ferromagnetic phase,
hence $S_{0B}=n_{0B}$.

We consider the propagation of the plane waves
which travel in the arbitrary direction $\textbf{k}=\{k_{x}, k_{y}, k_{z}\}$.
Hence, the perturbations have the following structure illustrated within the concentration
$\delta n=N\cdot e^{-\imath\omega t+\imath \textbf{k} \textbf{r}}$,
where $N$ is the constant amplitude.
Therefore, equations are transformed to the algebraic equations:
\begin{equation}\label{BECfinTSpin1} \omega\delta n_{B}=n_{0B} (\textbf{k}\delta \textbf{v}_{B}), \end{equation}
\begin{equation}\label{BECfinTSpin1} mn_{0B}\omega\delta \textbf{v}_{B}=n_{0B}\textbf{k}(g_{1}\delta n_{B}+g_{2}\delta S_{B}^{z}), \end{equation}
and
\begin{equation}\label{BECfinTSpin1} \omega\delta S_{B}^{z}=S_{0B}^{z}(\textbf{k}\delta \textbf{v}_{B}),\end{equation}
\emph{and} two other equations
\begin{equation}\label{BECfinTSpin1} -\imath\omega\delta S_{B}^{x}=(p-q)\delta S_{B}^{y}, \end{equation}
and
\begin{equation}\label{BECfinTSpin1} -\imath\omega\delta S_{B}^{y}=-(p-q)\delta S_{B}^{x}. \end{equation}

Set of hydrodynamic equations splits on two sets of equation.
One for $\delta n_{B}$, $\delta \textbf{v}_{B}$, and $\delta S_{B}^{z}$
which gives the acoustic wave spectrum:
\begin{equation}\label{BECfinTSpin1} \omega^{2}=(g_{1}+g_{2})n_{0}k^{2}/m . \end{equation}
Another set is for $\delta S_{B}^{x}$ and $\delta S_{B}^{y}$
which give the constant frequency in the quasi-classic limit:
\begin{equation}\label{BECfinTSpin1} \omega=\mid q-p\mid/\hbar. \end{equation}

Since the nematic tensor density is reduced to the concentration and the spin density
we do not consider the nematic tensor evolution equation.
However, the nematic tensor evolution equation provides additional wave solution for the partially spin polarized BECs,
where the deformation mode appears.
It is related to elements $N^{xx}$, $N^{yy}$, and $N^{xy}$.
Moreover, the nematic tensor evolution contributes in the dynamics of the spin wave via the evolution of elements $N^{xz}$ and $N^{yz}$ for the partially spin polarized BECs.




\subsection{Spectrum in two-fluid hydrodynamics of the BEC and normal fluid}

We consider the plane wave small amplitude perturbations of the equilibrium state of the ferromagnetic phase of the finite temperature spin-1 bosons.
The equilibrium state is characterized by two constant concentrations $n_{0B}$, $n_{0n}$, zero velocity fields $\textbf{v}_{0B}=0$, $\textbf{v}_{0n}=0$,
and the spin densities $\textbf{S}_{0B}=S_{0B} \textbf{e}_{z}$ and $\textbf{S}_{0n}=S_{0n} \textbf{e}_{z}$,
where $S_{0B}=n_{0B}$ and $S_{0n}=n_{0n}$.
Hence, we have the following set of equations for the perturbations:
\begin{equation}\label{BECfinTSpin1} \omega\delta n_{a}=n_{0a} (\textbf{k}\delta \textbf{v}_{a}), \end{equation}
$$mn_{0B}\omega\delta \textbf{v}_{B}=n_{0B}\textbf{k}(g_{1}\delta n_{B}+g_{2}\delta S_{B}^{z})$$
\begin{equation}\label{BECfinTSpin1}
+2n_{0B}\textbf{k}(g_{1}\delta n_{n}+g_{2}\delta S_{n}^{z}), \end{equation}
$$mn_{0n}\omega\delta \textbf{v}_{n}=2n_{0n}\textbf{k}(g_{1}\delta n_{B}+g_{2}\delta S_{B}^{z})$$
\begin{equation}\label{BECfinTSpin1}
+2n_{0n}\textbf{k}(g_{1}\delta n_{n}+g_{2}\delta S_{n}^{z}), \end{equation}
and
\begin{equation}\label{BECfinTSpin1} \omega\delta S_{a}^{z}=S_{0a}^{z}(\textbf{k}\delta \textbf{v}_{a}),\end{equation}
\emph{and} for equations for the spin projection evolution
\begin{equation}\label{BECfinTSpin1} -\imath\omega\delta S_{a}^{x}=(p-q)\delta S_{a}^{y}, \end{equation}
and
\begin{equation}\label{BECfinTSpin1} -\imath\omega\delta S_{a}^{y}=-(p-q)\delta S_{a}^{x}. \end{equation}

The spin waves for the BEC and for the normal fluid have same frequency in the ferromagnetic phase:
\begin{equation}\label{BECfinTSpin1} \omega=\mid q-p\mid/\hbar. \end{equation}

We have the following set of equations for the perturbations of the concentrations of the BEC and the normal fluid
\begin{equation}\label{BECfinTSpin1} m\omega^{2}\delta n_{B}=n_{0B}k^{2}(g_{1}+g_{2})[\delta n_{B} +2\delta n_{n}], \end{equation}
and
\begin{equation}\label{BECfinTSpin1} m\omega^{2}\delta n_{n}=2n_{0n}k^{2}(g_{1}+g_{2})[\delta n_{B} +\delta n_{n}]. \end{equation}

Presence of the spin density projection in the direction of the equilibrium spin polarization $\delta S_{a}^{z}=\delta n_{a}$
changes the speed of two sounds via the shift of the interaction constant to $g_{1}+g_{2}$
in the same way as it is happens for the zero temperature.





\section{Conclusion}

The spin-1 BECs have been actively studied over 20 years.
Corresponding non-linear Schrodinger equations have been used in the large part of these studies.
This paper has been focused on the derivation of the well-known model from the microscopic motion of the quantum particles,
where the many-particle Schrodinger equation in the coordinate representation has been used as the starting point of the derivation.
Moreover, the generalization of the existing minimal coupling model has been made to include the contribution of the small temperatures.
Therefore, the system of spin-1 bosons in presented as the system of two quantum liquids
(more like two collective degrees of freedom of single species)
the BEC presenting particles in the quantum states with the lowest energy and the normal quantum fluid presenting particles being in the quantum states with energies above the minimal energy.

Hydrodynamic equations are derived for the arbitrary spin state of liquid.
However, the small temperature ferromagnetic phase and the polar phase are discussed as the limit regimes of the general model.

The set of hydrodynamic equations obtained from the Schrodinger equation leads to the set of $2\times N\times3$ functions
(scalar functions or the elements of tensors),
where $N$ is the number of particles,
the multiplier "2" comes from the fact that
spinless part of evolution of each particle is described by single complex function identical to two real functions,
the multiplier "3" comes from three projections of spin for each particle.
So, it is almost "infinite" number of equations.
Therefore, it should be truncated.

The step to make the truncation is made in accordance with the existing with the minimal coupling model
giving the non-linear Schrodinger equation with three component vector-spinor macroscopic wave function.
Therefore,the evolution of the concentration, velocity field, spin density, and the nematic tensor density are derived.
These equations are derived for the arbitrary temperature and the arbitrary interaction strength.
Next, the term describing the interaction are calculated
in the small temperature limit for the weakly interacting particles with the small range of action of the interaction potential.

Finally, each equation is separated on two parts:
one for the particles in the BEC state,
another for the normal fluid.
Hence, the two-fluid model of the small-temperature spin-1 bosons has been obtained.

These equations have been used to study the bulk small amplitude collective excitations in the form of the plane waves in the infinite medium.

\section{Acknowledgements}

Work is supported by the Russian Foundation for Basic Research (grant no. 20-02-00476).
This paper has been supported by the RUDN University Strategic Academic Leadership Program.

\section{Appendix: Hydrodynamic functions in terms of macroscopic wave functions}

During the derivation of the hydrodynamic equations from the microscopic model
we present the definitions of the hydrodynamic functions via the many-particle microscopic wave functions.
However, the introduction of the non-linear Schrodinger equations for the effective macroscopic wave function requires the presentation of the relation
of the hydrodynamic functions with this effective wave function.
The concentration and the spin density are presented above,
while other functions are given here in the appendix:
\begin{equation}\label{BECfinTSpin1} j_{a}^{\alpha}=\frac{1}{2m}(\hat{\Phi}_{a}^{\dag}\hat{p}^{\alpha}\hat{\Phi}_{a}+h.c.), \end{equation}
\begin{equation}\label{BECfinTSpin1} \Pi_{a}^{\alpha\beta}=\frac{1}{4m^{2}}(\hat{\Phi}_{a}^{\dag}\hat{p}^{\alpha}\hat{p}^{\beta}\hat{\Phi}_{a}
+(\hat{p}^{\alpha}\hat{\Phi})_{a}^{\dag}\hat{p}^{\beta}\hat{\Phi}_{a}+h.c.), \end{equation}
\begin{equation}\label{BECfinTSpin1} N_{a}^{\alpha\beta}=\frac{1}{2}\hat{\Phi}_{a}^{\dag}(\hat{F}^{\alpha}\hat{F}^{\beta}+\hat{F}^{\beta}\hat{F}^{\alpha})\hat{\Phi}_{a}, \end{equation}
\begin{equation}\label{BECfinTSpin1} J_{a}^{\alpha\beta}=\frac{1}{2m}(\hat{\Phi}_{a}^{\dag}\hat{F}^{\alpha}\hat{p}^{\beta}\hat{\Phi}_{a}+h.c.), \end{equation}
and
\begin{equation}\label{BECfinTSpin1} J_{N,a}^{\alpha\beta\gamma}
=\frac{1}{4m}(\hat{\Phi}_{a}^{\dag}(\hat{F}^{\alpha}\hat{F}^{\beta}+\hat{F}^{\beta}\hat{F}^{\alpha})\hat{p}^{\gamma}\hat{\Phi}_{a}+h.c.). \end{equation}

\end{document}